# Contra-Analysis: Prioritizing Meaningful Effect Size in Scientific Research

**Authors:** Bruce A. Corliss[1,*], Yaotian Wang[2], Heman Shakeri[1], Philip E. Bourne[1,3]

**Affiliations:**

[1]School of Data Science, University of Virginia; Charlottesville, Virginia

[2]Department of Statistics, University of Pittsburgh; Pittsburgh, Pennsylvania

[3]Department of Biomedical Engineering, University of Virginia; Charlottesville, Virginia

*Corresponding author. Email: bac7wj@virginia.edu

# Abstract

At every phase of scientific research, scientists must decide how to allocate limited resources to pursue the research inquiries with the greatest potential. This prioritization dictates which controlled interventions are studied, awarded funding, published, reproduced with repeated experiments, investigated in related contexts, and translated for societal use. There are many factors that influence this decision-making, but interventions with larger effect size are often favored because they exert the greatest influence on the system studied. To inform these decisions, scientists must compare effect size across studies with dissimilar experiment designs to identify the interventions with the largest effect. These studies are often only loosely related in nature, using experiments with a combination of different populations, conditions, timepoints, measurement techniques, and experiment models that measure the same phenomenon with a continuous variable. We name this assessment contra-analysis and propose to use credible intervals of the relative difference in means to compare effect size across studies in a meritocracy between competing interventions. We propose a data visualization, the contra plot, that allows scientists to score and rank effect size between studies that measure the same phenomenon, aid in determining an appropriate threshold for meaningful effect, and perform hypothesis tests to determine which interventions have meaningful effect size. We illustrate the use of contra plots with real biomedical research data. Contra-analysis promotes a practical interpretation of effect size and facilitates the prioritization of scientific research.

# Introduction

Scientific consensus is the evidence-based collective agreement of the scientific community to designate an assertion as scientific knowledge (*1*). While scientific consensus is considered a cornerstone of science, this process represents a minority of the decision-making in the lifecycle of scientific research. Scientific research is driven by research questions that are posed to expand scientific knowledge (*2*, *3*). Yet there is a near infinite number of research questions that can be formulated and studied. In contrast, the resources available for scientific research will always remain extremely limited relative to this near-infinite search space. There is limited funding available for scientific research and a limited number of personnel with limited time to pursue such endeavors (*4*). In practice, practitioners of the scientific method experience near constant decision-making for determining which research inquiries to pursue (*5*). Resources must be allocated to investigate the interventions with the most potential (*6*) (we use the term intervention loosely here to mean a specific mechanism or a general approach that induces change in the system studied (*7*)). There are many factors that influence the potential of an intervention that are often context specific. In biomedical research, such factors can include the intervention's effect size, feasibility for real-world use, the cost and risk of gaining regulatory approval, the risk of adverse side-effects, competing treatments either in use or under development, public opinion, and the potential market size (*8*).

Prioritizing interventions based on these practical, social, ethical, and societal considerations go outside the realm of the scientific method and into decision theory (*9*). Decision theory and decision-making have been researched extensively and applied across scientific fields, including the process of scientific research. There have been frameworks developed to bridge scientific knowledge and decision-making through a formal analysis of beliefs and values (*10*). Decision theory has also been developed for automated decision-making with large datasets (*11*). Most relevant to this work, an alternative framework to null hypothesis significance testing was proposed that considers statistical significance, effect size, opportunity cost, prior knowledge, and reproducibility into a single index of merit (*12*). This approach had limited adoption, perhaps from disagreement over how these factors should be weighed against one another and integrated into a singular measure. Another possible reason is that not all considerations, such as the societal and behavioral factors we mentioned, can be easily incorporated into such a model.

Instead of developing a unified automated decision-making process that considers every possible factor, we seek to facilitate scientific decision-making by developing a better method to compare effect size between interventions. In the real world, effect size is the most important factor to consider when evaluating the practical significance of an intervention (*13*). A successful intervention must cause a meaningful change to the system, otherwise there would be no purpose to introduce the intervention to begin with (*14*). In general, interventions that demonstrate a larger effect size are often favored because the strength of their effect can typically be reduced by adjusting the intervention or its method of implementation (e.g., reducing intervention intensity, frequency, or duration). For biomedical treatments, examples of such adjustments include the dosage of the drug, it's delivery method, and its frequency of administration. An intervention with a stronger effect is potentially useful for a larger variety of applications that require a range of different effect sizes (*13*). For example, a treatment that has a large effect size with reducing inflammation could be useful for acute applications that require a strong short-term reduction in inflammation (*15*) such as minimizing tissue damage during heart failure. The same treatment at reduced dosage may also be useful for chronic applications requiring a smaller reduction in inflammation, such as preventing long-term tissue damage in autoimmune disease (*16*).

We propose contra plots as a data visualization that can compare effect size across broadly related experiments that measure the same phenomenon. Scientists can quickly gain perspective of how effect size magnitude compares across related studies to inform their choice of an appropriate minimum threshold for meaningful effect. Scientists can then quickly identify which interventions have a meaningful effect with a hypothesis test done in real-time and without recalculation regardless of the value of the threshold chosen. In addition, the visualization allows scientists to score and rank the strength of effect size across studies. We use real data to illustrate the use of contra plots for a continuous variable using the relative difference in means as a measure of effect size. This streamlined process will aid scientists in determining which interventions should be prioritized for further study, awarded funding, published, reproduced with repeated experiments, investigated in related contexts, and translated for societal use.

## Problem Description

Prioritizing interventions by their effect size is a data-informed instance of decision-making. Yet performing this task is challenging because there is no statistical framework to do so. Meta-analysis and similar approaches combine results across studies testing the same intervention that essentially represent repeated experiments (*17*). Results between sufficiently similar experiments can be pooled because a single underlying population effect size is assumed. In contrast, this prioritization process assumes different population effect sizes between different interventions. Studies are competing in a meritocracy rather than pooled together. Furthermore, studies considered for this prioritization will have a combination of different interventions, experiment designs, experiment models, populations, timepoints, measurement techniques, and measurement units (see Case Studies section for examples). There are simply too many different factors between each pair of experiments for an ANOVA analysis or similarly structured statistical approach that could rigorously compare effect size (*18*). Nonetheless, this prioritization of interventions must still be carried out, and there is pressure to do so effectively to optimally allocate scarce resources. *We define **contra-analysis** as the process of comparing effect size across different interventions that alter a common measured phenomenon under dissimilar experiment conditions*. A primary use of contra-analysis is to support scientific decision-making by identifying which interventions have evidence of a meaningfully large effect size and prioritize the candidates within this group.

To illustrate the use cases for this analysis technique, we outline twenty questions that scientists and peers encounter in the research lifecycle where contra-analysis could be used to inform the course of scientific research (Fig. 1). Gauging the potential of an intervention by its effect size is crucial for all these questions and their associated decisions. As an example- an individual scientist must continuously determine whether an intervention is worth pursuing further at each stage of scientific research, from ideation and data collection to publishing and translation. This assessment is crucial because scientists must overcome the bias of the sunk cost fallacy (*19*) to avoid wasting resources on an intervention that does not provide the effect size that was hoped for. Scientific peers must also determine if the potential effect size reported in a grant proposal is sufficiently large to merit grant funding or to match journal prestige during peer review. Additionally, peers must decide whether to collaborate on a given intervention based on its potential for a meaningful effect size.

While each of these questions have distinct considerations that is determined by a variety of factors, we propose three tasks focused on effect size that are shared across these decision-making events:

1. **Informed threshold selection**: an intervention should only be pursued if there is evidence that its effect size is large enough to be meaningful for practical purposes. While there is no standard approach for determining the minimum threshold for meaningful effect size, the value chosen should consider precedence established from previous research. To consider previous results, a method is required to visualize and compare effect size between broadly related experiments.

2. **Hypothesis testing for meaningful effect**: interventions with evidence of meaningful effect size should be prioritized for further development. Scientists need to determine which experiments have a meaningful effect size. This designation could be accomplished with a hypothesis test that determines if the effect size exceeds the minimum threshold.

3. **Effect size scoring**: there could remain a broad range of effect size magnitudes among the studies determined to have meaningful effect. It would be useful to score and rank the effect sizes to highlight and prioritize the most promising interventions within this subgroup.

## Contra-Analysis and Contra Plots

We propose contra plots as a data visualization that can be used to complete the effect size related tasks required to answer the questions posed in Fig. 1. Contra plots can be used to summarize the results from broadly related experiments that measure the same phenomenon. This data visualization is designed as a central step in the process of contra-analysis. In contra-analysis, scientists carry out a continuous decision-making cycle that compares the effect size of competing interventions on a common measured phenomenon (Fig 2A). A central process in this cycle is the use of contra plots.

We define the contra plot as a data visualization composed of a plot and table juxtaposed horizontally, similar to how forest plots are structured (*20*). The plot section visualizes the interval estimates effect size from each experiment and the table section contains metadata about the experiments the estimates were derived from (Fig 2B). The interval estimates have a multiple comparisons correction that matches the design of each study individually. Following the practice of previous studies (*21–24*), we use a Bonferroni correction for the credible interval estimates in this work. We focus on experiments that measure a continuous variable, so the relative difference in means is used to measure effect size for this case. The interval estimates are horizontally displayed and aligned by row with metadata from the experiment they belong to. The contents of the metadata are flexible and should be tailored to maximize interpretability. The metadata would at least include a description of the intervention used and the originating study, but could also include information about the species, control group, experiment model, timepoints, dosages, disease model, and population studied. We strongly recommend providing supplemental metadata in a separate table so that scientists can look-up additional information if necessary (see examples in Tables S1, S2).

In addition to the experiment metadata, there is also a single statistic that is calculated that is annotated as "Ls%" (which denotes the least percent, abbreviated to minimize horizontal space on the table). This statistic is simply the minimum value within each interval estimate for the relative difference in means. For interval estimates that include zero, Ls% is zero, while for entirely nonzero interval estimates Ls% is simple the closest interval bound to zero (See Methods: Scoring and Ranking Effect Sizes and Supplementary Methods for more details and a mathematical explanation).

We provide a brief overview of contra-analysis and how to use contra plots. A more detailed explanation of this process is found in the Methods section (see Bayesian Summary of Difference in

Means, Interval Estimation, Hypothesis Testing), and a mathematical justification is found in Supplementary Methods.

1. **Compile results and produce contra plot**: to start this process, scientists compile results that measure the same phenomenon. These results should be from studies that have passed peer review. The scientist will record summary statistics of the data into a CSV file such as the mean, standard deviation, and sample size of the control group and experiment group (see supplementary files 1 and 2 for examples). These can be directly copied from the text, obtained from correspondence with the authors, or estimated from the plots that visualize the data. The scientist will also record various information about the study for the metadata table in the contra plot and any supporting supplementary tables. Each study is assigned a unique identifier number so scientists can look-up additional information in the supplementary metadata table. Once related results are collected, scientists will run the contra plot function to generate the data visualization (Fig 2B).

2. **Determine threshold for meaningful effect**: the scientist determines their minimum threshold for a meaningful effect size. Since this threshold represents a minimum, an interval estimate with meaningful effect must entirely exceed this threshold. While scientists can specify whatever threshold they wish, they need an evidence-based justification for its value. Contra plots facilitate this discussion with a straightforward visualization of effect size. An appropriate threshold would ideally pass highly regarded past studies that reported a positive result. A threshold must also balance those results against the requirements of possible real-world applications. Threshold choice could also give priority to studies with results measured from the final model system (e.g., the human body in biomedical sciences, with results reported from clinical records or from clinical trials). Justification for the chosen threshold should be explicitly discussed in the text describing the result.

3. **Identify interventions with meaningful effect**: scientists perform a hypothesis test to determine which studies have meaningful effect size based on the specified threshold (Fig 2D). For a positive signed threshold, the scientist will visually check which studies have their entire interval estimate greater than the threshold. For a negative signed threshold, the scientist will visually check which studies have their entire interval estimate less than the threshold.

4. **Score /rank interventions by effect size**: the studies displayed in contra plots are sorted by the minimum effect size contained within their interval estimates (Fig 2E). This minimum value is either the closest interval bound to the origin (or zero if the interval estimate encloses zero). With this ranking, all studies that exceed the threshold will either be found at the top or bottom of a contra plot. The strength of effect for each study is also scored by this same value. The scores are listed in the "Ls%" column of the metadata table (see Fig 3A for example). This scoring will allow for comparing effect size strength between studies that are determined to have meaningful effect size.

5. **Choose interventions to pursue**: a scientist will determine which interventions are worthy of further pursuit and development based on the results of the hypothesis test and effect strength scores, along with a plethora of practical, social, ethical, and societal considerations.

6. **Obtain new results**: resources will be allocated for pursuing the most promising interventions, and new experimental results will be obtained.

7. **Recompilation**: the new experimental results will be incorporated into the contra plot, and the summary and decision-making process of contra-analysis will repeat.

Contra plots could be shared as research output in review papers or presented as a dynamic web application that continuously adds new studies to the plot as they are published. For primary research papers, contra plots can be used to justify the claims for the impact practical significance(*14*) of reported effect sizes. For scientific peers, contra-analysis will influence which interventions get prioritized for funding and publication.

## Results

We compiled results from studies of atherosclerosis to illustrate how contra-analysis can be used to visualize effect size across broadly related studies (Fig 3, 4). Atherosclerosis is the underlying cause of approximately 50% of all deaths in developed nations (20) and is characterized by the build-up of fatty deposits, called plaques, on the inner walls of arteries. Researchers use dietary, behavioral, pharmacological, and genetic interventions to study atherosclerosis and measure various biological phenomenon to monitor disease severity, including plasma cholesterol and plaque size. We will illustrate how contra plots can be used to rank and score effect size across studies, inform the selection of a threshold for meaningful effect size, and determine which experiments meet this threshold. Interventions that have effect sizes that are highly ranked and exceed the threshold should be given extra consideration for further development, although a variety of factors beyond effect size will influence this decision.

**Case Study 1: Total Plasma Cholesterol**

Total plasma cholesterol (TPC) is an important clinical read out for evaluating patients with atherosclerosis. Lowering TPC is therapeutic in most cases (depending on the composition of the cholesterol (20), which is beyond the focus of this example). A multitude of studies have reported results that demonstrate various interventions lowering TPC. Scientists must collectively decide which interventions demonstrate the most potential for further research, funding, commercialization, and translation. TPC levels vary from 60-3000 mg/dL across animal models used to research atherosclerosis and are reported in units of mmol/L as well (Table S1). This large variation in the measurement values makes it necessary to evaluate effect size on a relative scale.

We generate a contra plot to visualize the interventions reported to reduce TPC. Interventions that had no discernable effect on TPC are also included for reference. Since treatments that reduce TPC have been used in the clinic, determining a minimum threshold for effect size should consider current gold standard treatments and previous research studies. Statins are considered a gold standard treatment and clinical results indicate a 27% - 60% reduction in cholesterol (25) (only the LDL-C subtype of cholesterol was measured for these results, but for the purpose of this case study, we are disregarding the complexities associated with subtypes). Referring to the source studies, a conservative estimate of TPC reduction would roughly be at least a 10% reduction for statins at lower doses (*26*). Referring to the contra-plot, the threshold of 10% also provides some degree of separation between null results and those with a nonzero effect size (Fig 3A). Inspecting the interval estimates will readily reveal the studies that have effect sizes below the threshold of -10% (studies, 10, 11, 13, 15, 16, 18, 17, ordered by increasing effect size score) and are designated as having evidence of a meaningful effect size for reducing TPC. We score effect strength by reporting the value within each interval estimate closest to zero. For the case of interventions that reduce TPC, this would be the upper interval bound (reported in the "Ls%" column of the metadata section of the contra plot). Using this method, we see there is a considerable 4-fold spread of effect size strengths within the collection of interventions that have meaningful effect at a -10% threshold. This highlights that not all interventions with meaningful effect are equal, and this scoring can be used to inform decision-making for resource allocation within this subgroup of studies. This scoring method has the additional benefit of heavily penalizing results whose closest bound is near zero (at the edge of statistical significance) because these results have a lower chance of being reproduced (*27*) (e.g., study 21 versus 11).

The same analysis can be done with interventions that increase TPC (Fig 3B). Rather than identify possible treatments for atherosclerosis, these studies could be used to identify which interventions can be used as experiment models for atherosclerosis progression. Setting a threshold should be informed by the effect size reported from previously established experiment models. Setting the threshold of at least a 50% increase for the relative difference in means would pass the results from studies using gold standard experiment models for atherosclerosis. For this case, all values within the interval estimate must be greater than the 50% threshold. Just as with scoring negative signed effect sizes, we score positive signed effect size by reporting the value within each interval estimate closest to zero. With a threshold of 50%, we identify the interventions that have evidence of a meaningful effect size for increasing TPC (studies 32, 34, 29, 30, and 35).

Critically, this analysis process could be repeated by any scientists using any threshold with a single static contra-plot. The interval estimates in contra plots are invariant to the value of the threshold.

This allows scientists to form their own opinion for an appropriate threshold and facilitates discussion and consensus between scientists. More importantly, a single contra plot can be applied to multiple related contexts. For example, a contra plot of TPC could aid in determining different thresholds for meaningful effect size for early-stage atherosclerosis (favoring a long-term preventative treatment with lower effect size) versus late-stage atherosclerosis (favoring short-term treatments with larger effect size and more leniency for adverse side effects). Different thresholds could also be selected for conditions related to atherosclerosis, such as Moenckeberg medial calcific sclerosis and arteriolosclerosis (*28*) (hardening and calcification of medium and small size arteries, respectively).

**Case Study 2: Plaque Size**

As a second example, a similar case study examines the practical significance of therapeutic interventions that reduce plaque area as a treatment for atherosclerosis (Fig. 4A). Similar to measuring total cholesterol, plaque size is measured across units that span orders of magnitude (Table S2). Using the same general approach as the TPC example, a threshold of at least a 20% reduction would reveal that studies 17, 16, 18, and 23 have a meaningful effect size. However, examining interventions that increase plaque size reveals a more complex situation than those highlighted in the previous case study. The interventions that increase plaque size can be separated into those with extremely large effect size (studies 19 and 27 with a 1000% increase) and those with smaller increases (studies 24, 28, 25, 21, and 26 with 5% to 60% increase). We use a 500% decrease for a threshold to illustrate this separation (designating studies 19 and 27 with meaningful effect). While the relative difference in means is comparable across these groups, domain knowledge reveals that they represent distinct applications. The large effect groups represent interventions that initiate atherosclerosis. Examining the supplement table (STable 2) for this plot reveals that the control group measurements of this subgroup are clustered close to zero since healthy specimens have minimal plaque. The small effect size group have control groups are already afflicted with atherosclerosis and have considerable amounts of plaque within their blood vessels. They represent interventions that are related to further development of plaques and a worsening of the disease state. While contra plots can compare the relative difference in means between these subgroups, in many situations a scientist would not compare the two since the interventions would be used for separate applications.

## Discussion

We have identified a collection of decision-making problems that require comparing effect size between experiments with dissimilar designs that measure the same phenomenon. This decision-making process, named contra-analysis, is ubiquitously found throughout the lifecycle of scientific research. We propose contra plots as a graphical display to aid in this decision-making process. Contra plots can compare the strength of effect between different interventions and experiment designs using interval estimates of effect size. We illustrate the use of contra plots with credible intervals of the relative difference in means for continuous variables. Using real data, we show how contra plots can inform the selection of a threshold for meaningful effect and determine which interventions exhibit meaningful effect size via a hypothesis test. Contra plots can be used to inform decision-making for which interventions are worth pursuing compared to alternatives. This data visualization can aid in evaluating research inquiries for grant funding, peer review, repeated experiments, and translation for societal use. Using contra plots, scientists can quickly gain a holistic and broad perspective of what effect sizes are meaningful in a specific context and make informed decisions prioritizing interventions.

There are several avenues of research that could further improve this prioritization process. Future work could explore the use of less conservative multiple-comparisons corrections (*29*, *30*) than the Bonferroni-corrected intervals used in this study. The data visualization could also be extended to alternative statistics for different variable types, such as using the risk ratio or odds ratio for binary variables (*31*). There may be instances where experiments are similar enough that their interval estimates can be pooled with procedures used in meta-analysis. Further research would propose and validate a process to combine similar experiments into a single contra plot interval estimate. Additionally, forest plots emphasize studies with larger sample size by scaling the marker size of the point estimate for each interval. While placing emphasis on larger sample size is less relevant with contra-analysis (since the collection of studies include different experiment designs and study populations), there may be alternative metrics to prioritize studies. In the biomedical sciences, studies conducted in the final experiment model (humans in vivo) could be annotated to emphasize the heightened importance of their interval estimates.

Furthermore, a related question to identifying which interventions have meaningful effect size is which interventions have negligible (near-zero) effect size for a given application. Contra-analysis could be extended to determine if an intervention has a negligibly small effect size using similar strategies to the approach presented here. This assessment would essentially allow scientists to highlight statistically insignificant results that have evidence of near-zero effect size and conclude that an intervention is approximately independent from the experiment's measured phenomenon. For biomedical sciences, such a determination could help to eliminate alternative explanations for how an intervention works (i.d., its mechanisms of action (*32*)) or determine if undesirable side effects of a treatment are minimal enough to be negligible. Such a process may also provide a solution to the File Drawer Problem, where statistically insignificant results are underreported because there is no effective standard to analyze them (*33*).

There remains a significant issue with the x-axis scaling of contra plots. While positive sign effect sizes are bound between 0 and infinity, negative sign effect sizes are bound between 0 and -1 because a percent decrease is limited to 100% (assuming all measurements are greater than 0). This makes the negative x-axis distorted because a 2-fold change is four times further from the origin than a ½ fold change. Using a log transform would produce symmetric spacing for positive and negative signed effect sizes but would no longer maintain linearity when comparing effect sizes of the same sign. Since comparing the relative magnitude of effect size between studies is an important task for contra plots, this linear visual encoding for position should be preserved (*34*). A new transform may be required to make effect sizes equidistant to the origin while still maintaining linearity when comparing effect sizes with the same sign. Finally, statistics could be developed to conduct contra-analysis in a more statistical formal manner and possibly provide an alternative to the gold standard of using p-values (*35*, *36*).

Optimizing this prioritization process of deciding which interventions to pursue is critically important because it dictates the course of scientific research throughout the research lifecycle. There is a ponderous opportunity cost every time a suboptimal decision is made, either with pursuing an intervention that lacks a meaningful effect size or failing to pursue one that does. Contra-analysis aims to enable better decision-making for the prioritization and development of scientific research.

**Author Contributions:**

Conceptualization, Investigation, Writing- Original draft, Visualization, Data Curation, Software: BAC.

Methodology, Formal Analysis: BAC, YW.

Writing- Reviewing and Editing: BAC, HS, YW, PEB.

Supervision: PEB.

**Acknowledgements:** we would like to thank Dr. Kristin Naegle, PhD (UVA Biomedical Engineering and Computer Science) for her invaluable feedback. This work was funded by PEB's endowment for the School of Data Science, University of Virginia.

**Declaration of Interests:** The authors declare no competing interests.

**Data and Material Availability**: code used to generate all data and figures is written in R and available at: https://github.com/bac7wj/contra.

# Figures

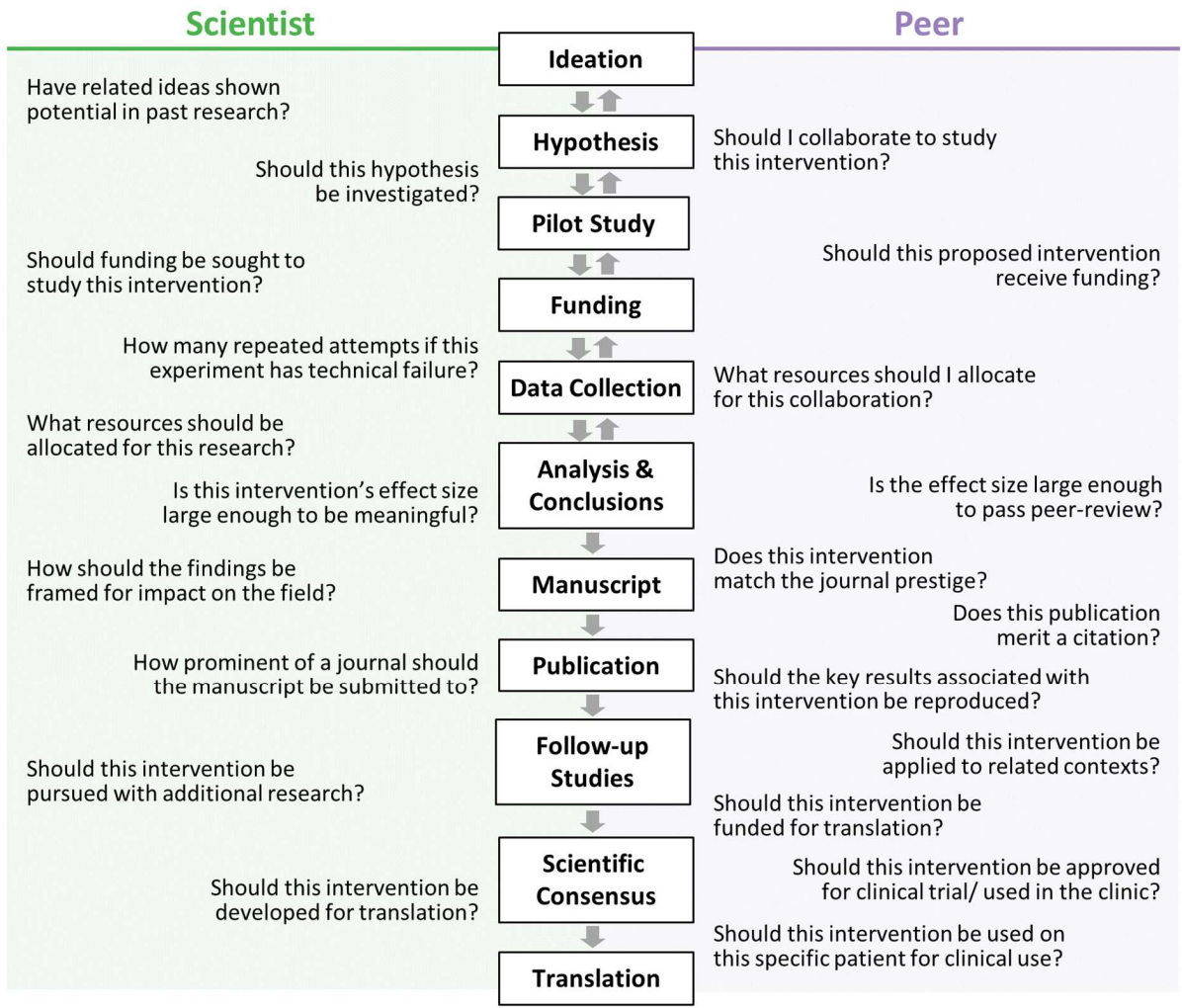

**Figure 1: Contra-analysis can inform critical decision-making steps in the scientific research lifecycle**. Flow chart of stages of the process of scientific research, from ideation to translation. For both the scientist pursuing the research (left column) and their peers (right column), a series of questions are listed that represent key decision-making steps that are informed by comparing effect size between different interventions that exert the same phenomenon of interest. Contra-analysis could be used to inform the decision-making to answer these questions (in conjunction with a plethora of practical, social, and philosophical considerations that must also be considered).

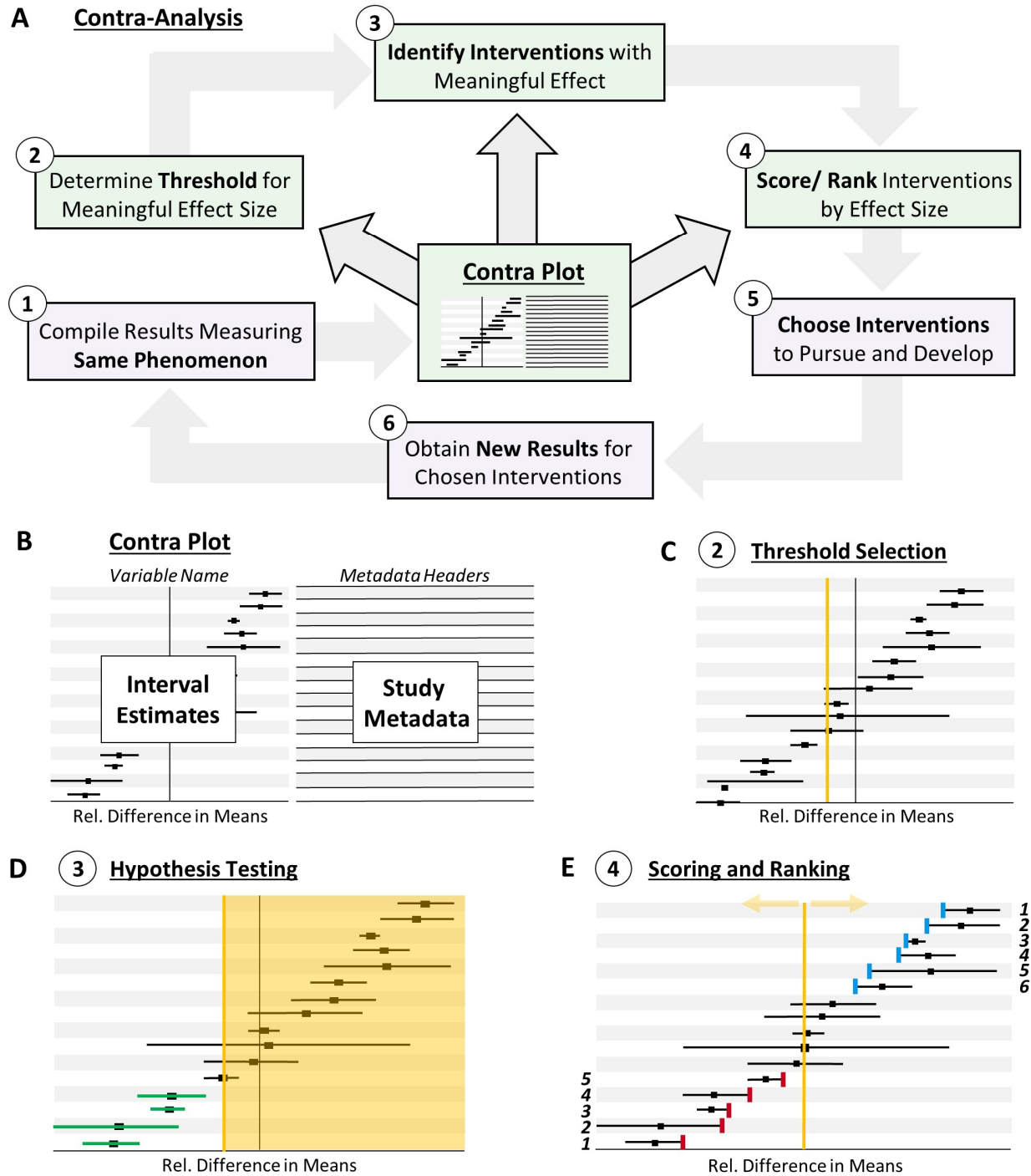

**Figure 2: Contra-analysis process flowchart and the uses of contra plots**. (**A**) Flow chart for contra-analysis, where scientists compare effect size across disparate experiments that measure the same phenomenon and decide which interventions are worthy of further pursuit and development. (**B**) Basic structure of a contra plot that includes a chart of interval estimates of effect size paired with a table of metadata describing the originating study for each result. (**C**) Visualization of threshold selection using contra plots. (**D**) Illustration of hypothesis testing with contra plots, where scientists determine which interventions have evidence of meaningful effect size. (**E**) Visualization of how effect size estimates are scored and ranked in contra plots to facilitate interpretation. Red vertical lines denote the credible bound that are used to score negative signed effect sizes. Blue vertical lines denote the credible bounds that are used to score positive signed effect sizes. These scores represent the smallest value for the threshold for meaningful effect size if it was shifted from the origin (gold vertical line and arrows). Italicized numbers at either side of plot are the rankings for interventions for negative and positive signed interventions based on the effect size scores.

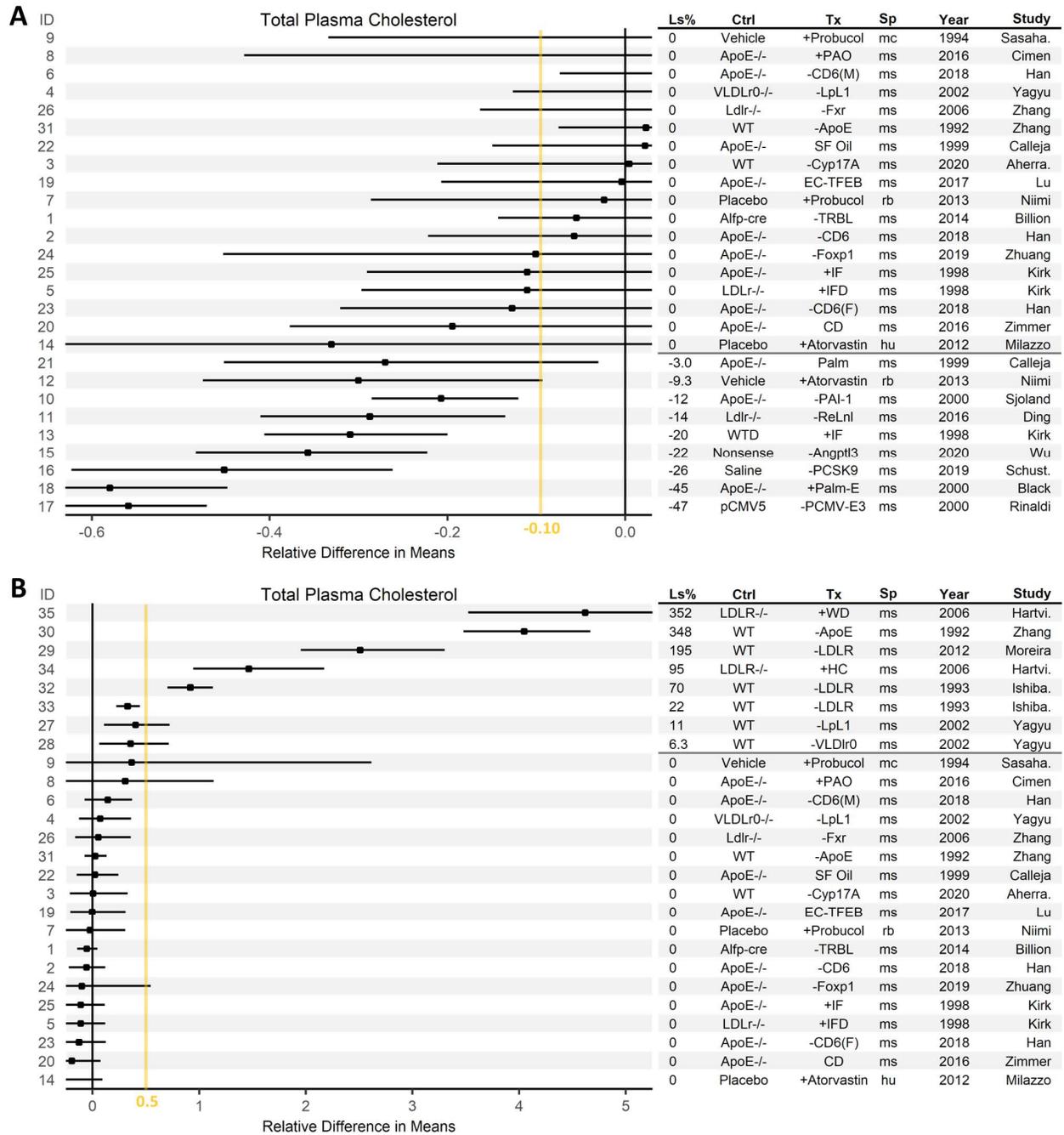

**Figure 3: Visualizing effect size across different experiments measuring total plasma cholesterol.**
**(A)** Contra plot of interventions that either reduced total plasma cholesterol or had no discernable effect (gold line delineates threshold for meaningful effect, included for illustrative purposes only and not formally part of the contra-plot). **(B)** Contra plot of interventions that either increased total plasma cholesterol or had no discernable effect. Abbreviations: Ls%, closest interval bound to zero, expressed as a percentage (the "least" value of the interval); Sp, species for experiment model; Ctrl, control group label; Tx, treatment label for experiment group. Intervals are 95% credible intervals of the relative difference in means, Bonferroni corrections are applied to design of each individual study (See STable 2) with no additional correction applied across studies included in contra plot.

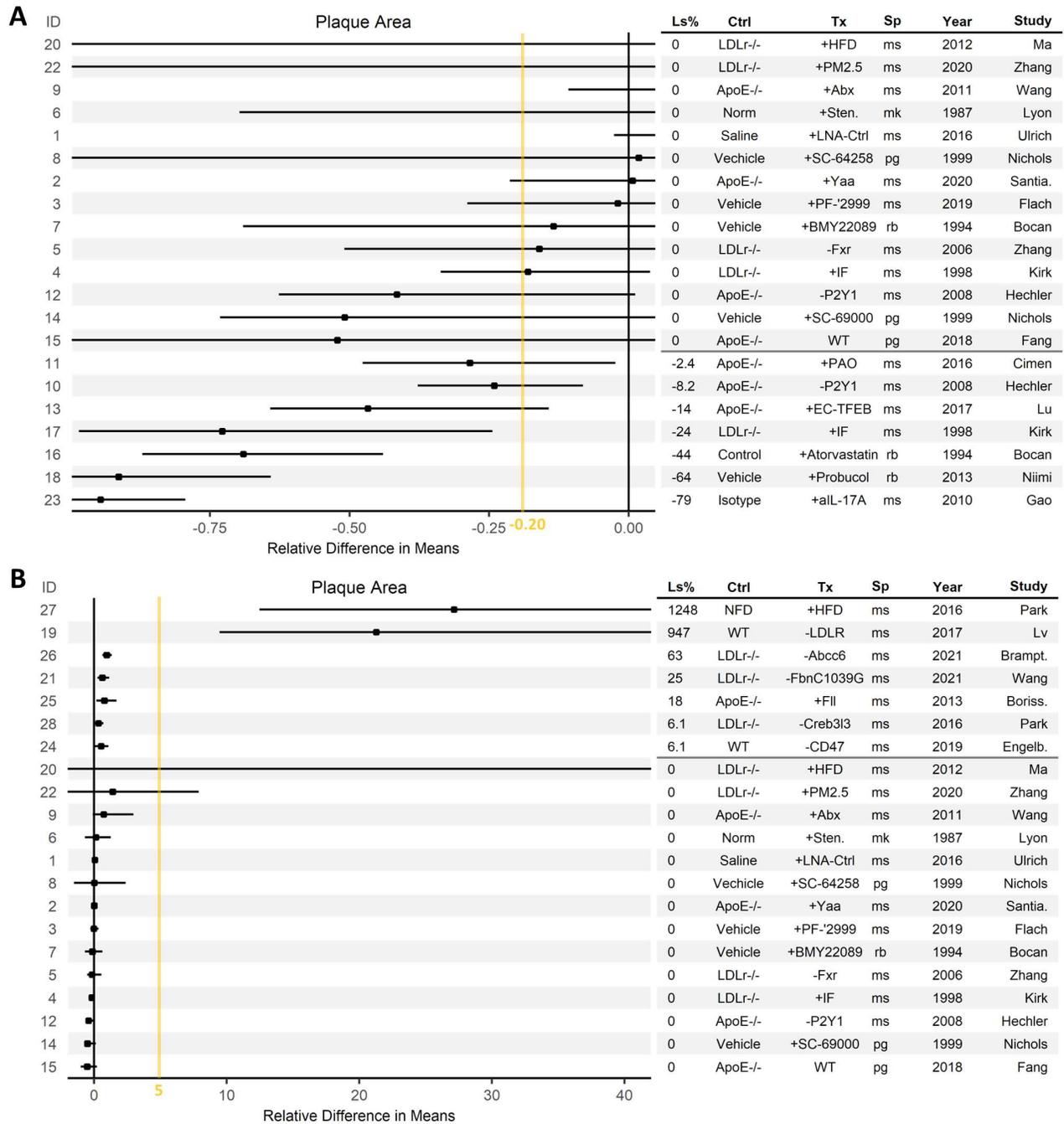

**Figure 4: Visualizing effect size across different experiments measuring total plasma cholesterol.**
**(A)** Contra plot of interventions that either reduced plaque area or had no discernable effect (gold line delineates threshold for meaningful effect, included for illustrative purposes only and not formally part of the contra-plot). **(B)** Contra plot of interventions that either increased plaque area or had no discernable effect. Abbreviations: Ls%, the value within the interval closest to zero; Sp, species for experiment model; Ctrl, control group label; Tx, treatment label for experiment group. Intervals are 95% credible intervals of the relative difference in means, Bonferroni corrections are applied to design of each individual study (See STable 2) with no additional correction applied across studies included in contra plot.

# Methods

**Code Availability**

All code used to generate all figures is written in R and available at: https://github.com/bac7wj/contra. Code was executed using RStudio 2022.07.0+548 "Spotted Wakerobin" Release for Windows (Windows NT 10.0; Win64; x64).

**Interval Estimation**

We draw our inspiration for this data visualization from forest plots in meta-analysis. Forest plots display the point estimate and 95% confidence interval of a specific statistic reported from a collection of studies examining the same intervention testing under similar experiment conditions (*31*). For the case of a continuous independent variable, forest plots often use the raw difference in sample means to find a better estimate for the true difference in population means (*31*) (alternative statistics are used for other types of variables, but are not the focus of this study). Unfortunately, using the raw difference in means for contra-analysis would not be effective because the means of the control and experiment group can vary orders of magnitude across different experiment designs, model systems, timepoints, and interventions. Raw differences cannot compare effect size across these datasets. However, the relative form of the difference in means can do so (difference in means divided by mean of control group). Other statistics of effect size, such as Cohen's D and related signal-to-noise measures, do not allow this assessment because normalizing the effect size by the variance loses any real world meaning.

There are several types of interval estimates that could theoretically provide a range of plausible values for the relative difference in means. Possible options include confidence intervals, credible intervals (*37*), and support intervals (*38*). However, not all these approaches can simply and consistently provide interval estimates for the relative difference in means, which represents a ratio of two normal variables. Although confidence intervals remain the standard in many areas of scientific research (*37*), they cannot always compute the interval estimates for the ratio of two normal random variables (*39*) (see discussion in Supplementary Methods: Interval Estimation). Support intervals for the ratio of two normal variables have not yet been defined (*38*).

We use credible intervals because they can reliably and simply provide interval estimates for the ratio of two normal random variables (see calculation in Supplementary Methods: Interval Estimation). Credible intervals define a range that an unobserved parameter value falls within at a particular probability (i.e., a 95% credible interval defines a range that contains the parameter with 95% probability). This contrasts with confidence intervals that define a range that contain the true population parameter if the interval were calculated on repeated samples (i.e., a 95% confidence interval defines a region that contains the true parameter 95% of the time if the experiment was repeated). We calculate and visualize the 95% credible interval for estimating the relative difference in means. For studies with experiments with several study groups, we perform a Bonferroni multiple comparison correction to control for Type 1 error.

Credible intervals have the option for incorporating prior information by specifying the prior distribution. This distribution represents what the scientist initially believes the underlying distribution to be prior to analyzing the data, and can alter the bounds of the credible interval (*40*, *41*). For contra-analysis, we specifically exclude the use of unique priors for the credible interval from each study because we wish to summarize the data and data alone. This analysis is meant to be meritocracy of effect size between datasets and not the strength of prior that a scientist chooses to use or how much prior information is known (*42*). Instead, a noninformative uniform prior is used for all interval estimates in the contra plot (see Supplementary Methods: Bayesian Summary of Difference in Means).

**Threshold Selection**

While scientists can specify whatever threshold they wish, they still need an evidence-based justification for its value. The justification would consider the effect size required for the intervention's real-world application balanced with the effect sizes reported and thresholds specified in previously published studies with related results. Although we will not recommend an exact process how to balance these considerations, setting an initial threshold value would ideally provide some degree of separation between the interval estimates for previously reported results that are statistically significance and statistically insignificant (Fig 2C). Threshold selection will also require balancing the demands for an effect size large enough to bring meaningful change versus considering what effect sizes are feasible from past studies. This separation could also give priority to studies with experiments using the final model system (e.g. in biomedical sciences, studies with human subjects). An appropriate initial threshold would ideally pass highly regarded studies that report a positive result, although there may be justification to specify a higher threshold. Changing the initial value of this threshold should then be supported with concrete arguments that are explicitly discussed in the manuscript.

It is important to note that the sign for the threshold is important. Interventions with positive or negative effect sizes have different applications and require distinct thresholds. In biomedical sciences, typically one direction of effect is used to model a disease or condition, while the opposite direction is associated with potential treatments to reverse the change (*43*). Since the sign of effect size represents distinct domain-specific applications, should not be compared to each other. For example, when identifying treatments for patients with high blood pressure, we are only interested in interventions that reduce blood pressure. Interventions that increase blood pressure may be useful for other applications such as treating low blood pressure or for creating animal models of high blood pressure. As a result, we use separate contra plots to visualize positive and negative sign effect sizes.

**Hypothesis Testing**

Once a threshold is specified, a scientist can identify which interventions have evidence of a meaningful threshold by performing an interval-based hypothesis test. Such a test can be done on-the-fly by checking if all values within the credible interval are further from zero than the threshold in the specified direction (see Supplemental Methods: Hypothesis Testing for details). This test would check that the effect size is at least as large as the threshold specified. Although interval estimation is typically used for parameter estimation, hypothesis testing can be performed with credible intervals. The procedure checks if the threshold value is within the bounds of the interval (*44*) (Fig 2D).

An advantage of using intervals for hypothesis testing is that a scientists can perform the test using any threshold without requiring recalculation of a statistic or interval. This property of interval estimates allows the same contra plot to be used for hypothesis testing regardless of the threshold value specified. Therefore, scientists with different perspectives or studying different applications can use the same contra plot to perform this analysis. Interventions that exceed the threshold would be favorably considered for further study and development.

**Scoring and Ranking Effect Sizes**

While hypothesis testing will reveal which results have evidence of meaningful effect size and highlight options with potential for further development, it would be useful to compare effect size between studies already deemed meaningful. It is difficult make such comparisons with credible intervals because the location of the interval must be considered as well as the width. One way to rank results is the point estimate for the relative difference in means, but this disregards the uncertainty of the measure and the potential risk that the published result is a false positive. An example of this is with study 14 in Fig. 3A, where the point estimate of a -33% difference in means has an impressive ranking, but the high uncertainty of the credible interval suggests that the difference in means could plausibly have effect sizes that are much closer to zero than the point estimate. We propose to score the strength of results with the threshold that transitions from being outside of the interval versus inside it. This value is equal to the credible bound closest to zero, and represents the smallest effect size out of the set of plausible effect size contained within the interval (Fig 2D). Results are scored and ranked based on the magnitude of the closest credible bound to zero. This ranking scheme would also match the reverse order of which interval estimates would fail to exceed the threshold if the value was swept from zero. This approach provides a uniform standard in which to score and rank the potential of results from disparate experiments based on the interval estimate of the relative difference in means.

Note that our approach to summarizing effect strength follows the standard conventions for summarizing effect size. With the standard convention, effect size and direction are only reported when results are statistically significant (*45*) (i.e., when the credible interval does not include zero). We similarly only report a nonzero size of effect size if the credible interval does not contain zero, where we can make a Bayesian "claim with confidence" for a nonzero effect size (*45*) (we refer to this as Bayesian posterior significance (*46*)). Furthermore, results from a single study is deemed unreliable (*47*), even if the effect size is deemed statistically significant with a statistical test. Our approach of scoring and ranking effect sizes by their smallest plausible value within the interval estimate enforces a conservative interpretation of the data.

# Supplementary Materials for

# Contra-Analysis: Prioritizing Meaningful Effect Size in Scientific Research

**Authors:** Bruce A. Corliss[1,*], Yaotian Wang[2], Heman Shakeri[1], Philip E. Bourne[1,3]

**Affiliations:**

[1]School of Data Science, University of Virginia; Charlottesville, Virginia

[2]Department of Statistics, University of Pittsburgh; Pittsburgh, Pennsylvania

[3]Department of Biomedical Engineering, University of Virginia; Charlottesville, Virginia

*Corresponding author. Email: bac7wj@virginia.edu

**This PDF file includes:**

Supplementary Materials and Methods

Tables S1 to S2

Supplementary References

# Supplementary Methods and Materials

## Study Identification

The studies presented in in Fig. 3-4 and STable 1-2 were compiled based on a literature search using Pubmed, Google Scholar, and Google search. Included results were limited to papers that were indexed on Pubmed. Papers were identified based on searches with combinations of the following keywords.

Total cholesterol example: atherosclerosis, total cholesterol, cholesterol, plasma cholesterol, reduce, protect, increase, independent, no change, mouse, rabbit, human, primate, rat.

Plaque size example: plaque size, plaque area, lesion size, lesion area, reduce, protect, increase, independent, no change, mouse, rabbit, human, primate, rat.

The included results are not meant to be complete, but rather give the reader a simplified toy example with how the proposed metric could be used to ascertain the practical insignificance of results. The mean and standard deviation of each group were either copied directly from the source publication or estimated from the figure using Web Plot Digitizer (https://automeris.io/WebPlotDigitizer/).

## Bayesian Summary of Difference in Means

Let $X_1, ..., X_m$ be an i.i.d. sample from a control group with a distribution Normal($\mu_X, \sigma_X^2$), and $Y_1, ..., Y_n$ be an i.i.d. sample from an experiment group with a distribution Normal($\mu_Y, \sigma_Y^2$). Both samples are independent from one another, and we conservatively assume unequal variance, i.e., $\sigma_X^2 \neq \sigma_Y^2$ (the Behrens-Fisher problem (*1*) for the means of normal distributions).

We analyze data in a Bayesian manner using minimal assumptions and therefore use a noninformative prior, specified as

$$p(\mu_X, \mu_Y, \sigma_X^2, \sigma_Y^2) \propto (\sigma_X^2)^{-1} (\sigma_Y^2)^{-1}. \tag{1}$$

The model has a closed-form posterior distribution. We define the sample means $\bar{x}$ and $\bar{y}$, along with variances $s_X^2$ and $s_Y^2$ for the control group and experiment group. Specifically, the population means,

conditional on the variance parameters and the data, follow normal distributions:

$$\mu_X | \sigma_X^2, x_{1:m} \sim \text{Normal}\left(\bar{x}, \frac{\sigma_X^2}{m}\right) \text{ and} \quad (2)$$

$$\mu_Y | \sigma_Y^2, y_{1:n} \sim \text{Normal}\left(\bar{y}, \frac{\sigma_Y^2}{n}\right). \quad (3)$$

The population variances each independently follow an inverse gamma distribution (InvGamma):

$$\sigma_X^2 | x_{1:m} \sim \text{InvGamma}\left(\frac{m-1}{2}, \frac{(m-1)s_X^2}{2}\right) \text{ and} \quad (4)$$

$$\sigma_Y^2 | y_{1:n} \sim \text{InvGamma}\left(\frac{n-1}{2}, \frac{(n-1)s_Y^2}{2}\right). \quad (5)$$

We exclude the use of prior information in this analysis because we wish to summarize the data alone and not be influenced by the beliefs of the scientist reporting the data (specifically, the strength of the prior used can considerably influence the outputs of a Bayesian statistical analysis (*2, 3*)).

**Interval Estimation**

We wish to calculate an interval estimate for the relative difference in means $r\mu_{DM}$, defined as

$$r\mu_{DM} = \frac{\mu_Y - \mu_X}{\mu_X}. \quad (6)$$

We assume that $\mu_X > 0$ and $\mu_Y > 0$ since calculating percent change from measurements with mixed sign is nonsensical. The frequentist confidence interval for $r\mu_{DM}$ can be calculated with Filler's theorem (*4*). However, if the control sample mean is stochastically close to zero, one or both endpoints will not exist for this interval (*5*). For the sake of consistency and simplicity, we wish to summarize the data with interval endpoints that exist for all cases.

Fortunately, calculating the credible interval for $r\mu_{DM}$ does not have the same limitations as the confidence interval because the endpoints will exist regardless of the distance of the sample mean to zero. We first define a bounded interval that represents the lower and upper two-tailed credible bounds for $r\mu_{DM}$ and encompasses its plausible values. If $Q_{relative}(p)$ is the quantile function of this posterior at a given probability $p$, the bounds $c_{lo}$ and $c_{hi}$ satisfies

$$c_{lo} = Q_{relative}(\alpha_{DM}/2) \qquad (7)$$

$$c_{hi} = Q_{relative}(1 - \alpha_{DM}/2). \qquad (8)$$

However, this quantile function is difficult to compute, and there is no closed-form posterior distribution for $(\mu_Y - \mu_X)/\mu_X$. We define $F_{relative}(x)$ as the empirical cumulative distribution function (ECDF) of $r\mu_{DM}$ from $K$ Monte Carlo simulations (9). With $K$ samples from the posterior distribution of $\mu_y$ and $u_x$, the cumulative distribution function $F_{relative}$ is defined as

$$F_{relative}(z) = K^{-1} \sum_{i=1}^{K} \mathbb{I}\left(\frac{\mu_Y^i - \mu_X^i}{\mu_X^i} \leq z\right). \qquad (9)$$

We estimate the interval bounds by numerically solving for $\hat{c}_{lo}$ and $\hat{c}_{hi}$ such that

$$F_{relative}(\hat{c}_{lo}) = \alpha_{DM}/2, \qquad (10)$$

$$F_{relative}(\hat{c}_{hi}) = 1 - \alpha_{DM}/2. \qquad (11)$$

The interval $[\hat{c}_{lo}, \hat{c}_{hi}]$ represents the credible interval for $r\mu_{DM}$.

**Hypothesis Testing**

To determine if a result has a meaningful effect size, scientists can perform a hypothesis test to determine if $r\mu_{DM}$ is further from zero than a specified threshold in a particular direction. Our approach allows scientists to perform hypothesis tests visually by examining the credible intervals displayed in contra-plots without any calculations required. Interval estimates can be used for hypothesis testing against any threshold without recalculation of the interval. The procedure checks if the threshold (value of null hypothesis) is within the bound of the interval. Credible intervals have been used for hypothesis tests against any threshold at the same credible level as the interval (6–9). We note that there is controversy with using credible intervals for hypothesis testing because the size of the effect is estimated under the assumption that it is present (10). We perceive this reservation to be a nonissue because an effect size of zero has never been shown to exist in the real world and can't be confirmed with finite data (11). We assume a non-zero effect size is present in all cases, the question this procedure answers is whether there is evidence that it is large enough to be considered meaningful.

We approximately perform a composite hypothesis test that the absolute relative difference in means is greater than a threshold δ, specifically

$$H_0: |r\mu_{DM}| \leq \delta \; ; \; H_1: |r\mu_{DM}| > \delta. \qquad (12)$$

We use the magnitude of the value within [$\hat{c}_{lo}$, $\hat{c}_{hi}$] that is closest to zero as a test statistic to perform a hypothesis test. The test is conducted at the same credible level (1–$\alpha_{DM}$) as used to calculate the interval. We define this test statistics as |$\delta_L$|, where

$$|\delta_L| = \min_{z \in [\hat{c}_{lo}, \hat{c}_{hi}]} |z|. \qquad (13)$$

We would reject $H_0$ if and only if |$\delta_L$| > δ or |$\delta_L$| < –δ.

However, interventions with positive signed effect sizes have different applications than those with negative sign, and the threshold should be specific to each. Results with different signed effect sizes should not be compared to each other as implied with this hypothesis test. For example, when identifying treatments for patients with high blood pressure, we are only interested in interventions that reduce blood pressure. Interventions that increase blood pressure may be useful for other applications such as patients with low blood pressure or for creating animal models of high blood pressure. As a result, the specified thresholds for positive signed and negative signed effect sizes have distinct considerations and may differ greatly.

We instead perform a composite hypothesis test that uses a positive threshold for measuring the effect strength of positive signed effect sizes and a negative threshold for negative signed effect sizes. For a positive signed effect size, we perform a hypothesis test using the threshold $\delta^+$ (where $\delta^+ > 0$) of the form

$$H_0^+: r\mu_{DM} \leq \delta^+ \; ; \; H_1^+: r\mu_{DM} > \delta^+. \qquad (14)$$

We defined a test statistic that is the value within the interval estimate that is closest to zero (essentially the signed version of |$\delta_L$|), denoted as $\delta_L$

$$\delta_L = sign\left(\frac{\hat{c}_{lo} + \hat{c}_{hi}}{2}\right) \min_{z \in [\hat{c}_{lo}, \hat{c}_{hi}]} |z|. \qquad (15)$$

We reject $H_0^+$ and conclude practical significance if $\delta_L > \delta^+$ because $\delta_L$ is the posterior lower bound for $r\mu_{DM}$ for positive-signed effect sizes (or zero if the lower bound is less than zero).

For a negative signed effect size, we perform a hypothesis test using the threshold $\delta^-$ (where $\delta^- < 0$) of the form

$$H_0^-: r\mu_{DM} \geq \delta^- \;;\; H_1^-: r\mu_{DM} < \delta^- . \tag{16}$$

We reject $H_0^-$ and conclude practical significance if $\delta_L < \delta^-$ because $\delta_L$ is posterior upper bound for $r\mu_{DM}$ for negative-signed effect sizes (or zero if the upper bound is greater than zero). Note that this composite hypothesis test does not support testing if $r\mu_{DM}=0$ for the null hypothesis. The thresholds for $\delta^\pm$ must have a nonzero value because a point hypothesis is not supported for hypothesis testing within the Bayesian framework with credible intervals (*12*).

# Supplementary Tables

Supplementary Table 1: Interventions that change total plasma cholesterol.

| ID | Study | Year | Group X | $\bar{x}$ | $s_x$ | $n_x$ | Group Y | $\bar{y}$ | $s_y$ | $n_y$ | Units | $\alpha_{DM}$ | Sp | PMID | Loc | Sgn |
|---|---|---|---|---|---|---|---|---|---|---|---|---|---|---|---|---|
| 1 | Billion | 2014 | Alfp-cre | 3.45 | 0.24 | 6 | Alfp-creTR?fl/fl | 3.26 | 0.22 | 6 | mmol/L | 0.05 | ms | 24797634 | F1C | 0 |
| 2 | Han | 2018 | ApoE-/-CD6WT | 1251 | 161 | 10 | ApoE-/-CD6-/- | 1179 | 143 | 5 | mg/dl | 0.05 | ms | 29615096 | T1 | 0 |
| 3 | Aherra. | 2020 | Cyp17AWT (F) | 2.29 | 0.53 | 7 | Cyp17A-/- (F) | 2.3 | 0.32 | 6 | mmol/L | 0.05 | ms | 32472014 | T2 | 0 |
| 4 | Yagyu | 2002 | VLDLr0-/- LpL1WT | 141 | 34 | 8 | VLDLr0-/- LpL1-/- | 151 | 24 | 13 | mg/dl | 0.05 | ms | 11790777 | T1 | 0 |
| 5 | Kirk | 1998 | LDLr-/- | 105 | 28 | 15 | LDLr-/- + IFD | 93.4 | 29 | 15 | mmol/L | 0.05/3 | ms | 9614153 | T2 | 0 |
| 6 | Han | 2018 | ApoE-/-CD6WT | 2518 | 257 | 8 | ApoE-/-CD6-/- | 2876 | 506 | 6 | mg/dl | 0.05 | ms | 29615096 | T1 | 0 |
| 7 | Niimi | 2013 | Vehicle | 1335 | 269 | 8 | Probucol | 1303 | 376 | 8 | mg/dL | 0.05/12 | rb | 24188322 | F1 | 0 |
| 8 | Cimen | 2016 | ApoE-/- | 568 | 81 | 6 | ApoE-/- PAO | 742 | 256 | 4 | mg/dL | 0.05 | ms | 27683551 | F5E | 0 |
| 9 | asahar. | 1994 | Control | 11.2 | 8.78 | 8 | Probucol | 15.3 | 6.55 | 8 | mg/dL | 0.05 | mc | 8040256 | T1 | 0 |
| 10 | Sjoland | 2000 | ApoE-/- PAI-1WT | 2503 | 266 | 11 | ApoE-/- PAI-1-/- | 1984 | 252 | 13 | mg/dl | 0.05 | ms | 10712412 | T1 | -1 |
| 11 | Ding | 2016 | Ldlr-/-Ad-Gal-RelnFl | 2087 | 531 | 15 | Ldlr-/- Ad-Cre-RelnF | 1487 | 364 | 16 | mg/dl | 0.05 | ms | 26980442 | SF2B | -1 |
| 12 | Niimi | 2013 | Vehicle | 1335 | 269 | 8 | Atorvastin | 934 | 231.9 | 8 | mg/dL | 0.05/12 | rb | 24188322 | F1 | -1 |
| 13 | Kirk | 1998 | WTD -IF | 4.78 | 1.2 | 20 | WTD +IF | 3.3 | 0.67 | 20 | mmol/L | 0.05/4 | ms | 9614153 | F1A | -1 |
| 14 | Milazzo | 2012 | Placebo | 202 | 28.2 | 5 | 150 mg Atorvastatin | 135.2 | 64.2 | 5 | mg/dL | 0.05/6 | hu | 22716983 | ST9 | -1 |
| 15 | Wu | 2020 | Luciferase siSRNA | 238 | 15.7 | 5 | Angptl3 siRNA | 153 | 22.4 | 5 | mg/dL | 0.05/6 | ms | 32808882 | F1D | -1 |
| 16 | Schuste | 2019 | WTD +Saline | 463 | 103 | 12 | WTD +PCSK9-mAb1 | 254 | 108 | 10 | mg/dl | 0.05 | ms | 31366894 | F1A | -1 |
| 17 | Rinaldi | 2000 | pCMV5 | 642 | 63 | 9 | PCMV-E3 | 283 | 69 | 10 | mg/dl | 0.05 | ms | 11110410 | F3 | -1 |
| 18 | Black | 2000 | ApoE-/- | 300 | 97 | 13 | ApoE-/- +Palm-E | 126 | 41 | 14 | mg/dl | 0.05 | ms | 11015467 | T1 | -1 |
| 19 | Lu | 2017 | ApoE-/- | 969 | 219 | 6 | ApoE–/– EC-TFEB | 965 | 84.6 | 6 | mg/dL | 0.05 | ms | 28143903 | SF6A | 0 |
| 20 | Zimmer | 2016 | ApoE-/- | 956 | 181 | 5 | ApoE–/– + CD | 770 | 61.3 | 4 | mg/dL | 0.05 | ms | 27053774 | F2F | 0 |
| 21 | Calleja | 1999 | ApoE-/- | 444 | 109 | 9 | ApoE–/– + Palm Oil | 324 | 88 | 9 | mg/dL | 0.05 | ms | 10521366 | T2 | 0 |
| 22 | Calleja | 1999 | ApoE-/- | 585 | 131 | 11 | ApoE–/– + SF Oil | 598 | 102 | 11 | mg/dL | 0.05/7 | ms | 10521366 | T3 | 0 |
| 23 | Han | 2018 | WTD-KO-F | 2338 | 332 | 5 | WTD-DKO-F | 2040 | 474 | 10 | mg/dL | 0.05 | ms | 29615096 | T2 | 0 |
| 24 | Zhuang | 2019 | Foxp1WTApoE-/- | 1037 | 635 | 12 | Foxp1ECKOApoE–/– | 932.5 | 442 | 12 | mg/dL | 0.05 | ms | 31318658 | ST1 | 0 |
| 25 | Kirk | 1998 | LDLr-/- WTD -IF | 105 | 27.9 | 15 | LDLr–/– WTD +IF | 93.4 | 29.4 | 15 | mmol/L | 0.05/3 | ms | 9614153 | T2 | 0 |
| 26 | Zhang | 2006 | Ldlr-/- FxrWT (M) | 1888 | 627 | 13 | Ldlr–/– Fxr–/– (M) | 1988 | 462 | 13 | mg/dL | 0.05/2 | ms | 16825595 | F3B | 0 |
| 27 | Yagyu | 2002 | LpL1WT | 104 | 16 | 8 | LpL1–/– | 146 | 34 | 9 | mg/dl | 0.05 | ms | 11790777 | T1 | 0 |
| 28 | Yagyu | 2002 | VLDlr0WT | 104 | 16 | 8 | VLDlr0–/– | 141 | 34 | 8 | mg/dl | 0.05 | ms | 11790777 | T1 | 0 |
| 29 | Moreira | 2012 | C57Bl/6 | 55.6 | 15 | 12 | LDLR-/- | 195.1 | 15.6 | 9 | mg/dL | 0.05/2 | ms | 22810096 | F1A | 1 |
| 30 | Zhang | 1992 | C57Bl/6 | 86 | 20 | 46 | ApoE-/- | 434 | 129 | 40 | mg/dL | 0.05/3 | ms | 1411543 | T1 | 1 |
| 31 | Zhang | 1992 | C57Bl/6 | 86 | 20 | 46 | ApoE+/- | 88 | 22 | 47 | mg/dL | 0.05/3 | ms | 1411543 | T1 | 0 |
| 32 | Ishib. | 1993 | C57Bl/6 | 119 | 17 | 19 | LDLR-/- | 228 | 36 | 16 | mg/dL | 0.05/3 | ms | 8349823 | T1 | 1 |
| 33 | Ishib.i | 1993 | C57Bl/6 | 119 | 17 | 19 | LDLR+/- | 158 | 25 | 39 | mg/dL | 0.05/3 | ms | 8349823 | T1 | 0 |
| 34 | Hartv. | 2006 | LDLR-/- | 8.7 | 2.6 | 12 | LDLR-/- HC | 21.45 | 5.2 | 12 | mmol/L | 0.05/3 | ms | 17255537 | F2A | 1 |
| 35 | Hartv. | 2006 | LDLR-/- | 8.7 | 2.6 | 12 | LDLR-/- WD | 48.88 | 7.6 | 8 | mmol/L | 0.05/3 | ms | 17255537 | F2A | 1 |

*Abbreviations: ID, study identification number to compare studies between contra-plot and this supplemental table; Group X, control group label; $\bar{x}$, control group sample mean; $s_x$, control group sample standard deviation; $n_x$, control group sample size; group y, experiment group label; $\bar{y}$, experiment*

*group sample mean; $s_y$, experiment group sample standard deviation; $n_y$, experiment group sample size; Units, units of measure for variable of interest; $a_{DM}$, Bonferroni correction required for interval estimate; Sp, ; PMID, pubmed identification number of study; Loc, figure or table location of data in manuscript; Sgn, sign/ direction of effect size as reported from study (0 if result was reported as statistically insignificant).*

Supplementary Table 2: Interventions that change plaque size

| ID | Study | Year | Group X | $\bar{x}$ | $s_x$ | $n_x$ | Group Y | $\bar{y}$ | $s_y$ | $n_y$ | Units | $\alpha_{DM}$ | Sp | PMID | Loc | Sgn |
|---|---|---|---|---|---|---|---|---|---|---|---|---|---|---|---|---|
| 1 | Ulrich | 2016 | Saline | 48.9 | 3.3 | 9 | LNA-Control | 51.6 | 5.2 | 13 | % | 0.05/3 | ms | 27137489 | F1C | 0 |
| 2 | Santiago. | 2020 | ApoE-/- | 6.7E+05 | 1.1E+05 | 8 | ApoE-/-.Yaa | 6.7E+05 | 1.6E+05 | 8 | µm2 | 0.05 | ms | 33110193 | F1A | 0 |
| 3 | Flach | 2019 | Vehicle | 3.7E+05 | 1.3E+05 | 14 | PF-'2999 | 3.6E+05 | 1.6E+05 | 15 | µm2 | 0.05/1 | ms | 30889221 | F2B | 0 |
| 4 | Kirk | 1998 | LDLr-/- -IF | 360 | 94 | 9 | LDLr-/- +IF | 295 | 36 | 8 | mm2 | 0.05/1 | ms | 9614153 | F2A | 0 |
| 5 | Zhang | 2006 | LDLr-/- FxrWT | 23.1 | 11.3 | 7 | LDLr-/- Fxr-/- | 19.4 | 7.2 | 8 | % | 0.05/2 | ms | 16825595 | F2B | 0 |
| 6 | Lyon | 1987 | Prox. -Sten. | 36 | 23 | 15 | Prox. +Sten. | 41 | 50 | 13 | % | 0.05 | mk | 3795393 | T3 | 0 |
| 7 | Bocan | 1994 | Vehicle | 0.713 | 0.297 | 8 | BMY22089 | 0.617 | 0.45 | 8 | mm2 | 0.05/7 | rb | 7840808 | T4 | 0 |
| 8 | Nichols | 1999 | Vehicle | 304788 | 113425 | 4 | SC-64258 | 310284 | 160647 | 3 | µm2 | 0.05/3 | pg | 10571535 | T2 | 0 |
| 9 | Wang | 2011 | ApoE-/- -Abx | 9956 | 11578 | 20 | ApoE-/- +Abx | 17196 | 13373 | 18 | µm2 | 0.05/6 | ms | 21475195 | F5E | 0 |
| 10 | Hechler | 2008 | ApoE-/- | 0.54 | 0.12 | 16 | ApoE-/-P2Y1-/- | 0.41 | 0.1162 | 15 | mm2 | 0.05 | ms | 18663083 | F2B | -1 |
| 11 | Cimen | 2016 | ApoE-/- | 225000 | 72732 | 10 | ApoE-/- + PAO | 161000 | 47434 | 10 | µm2 | 0.05 | ms | 27683551 | F5C | -1 |
| 12 | Hechler | 2008 | ApoE-/- | 21.2 | 8.08332 | 6 | ApoE-/-P2Y1-/- | 12.4 | 2.9394 | 6 | % | 0.05 | ms | 18663083 | F1A | -1 |
| 13 | Lu | 2017 | ApoE-/- | 16.1 | 7.7 | 10 | ApoE-/-EC-TFEB | 8.58 | 3.3 | 12 | % | 0.05 | ms | 28143903 | F7F | -1 |
| 14 | Nichols | 1999 | Vehicle | 304788 | 113425 | 4 | SC-69000 | 149779 | 34576 | 7 | µm2 | 0.05/3 | pg | 10571535 | T2 | -1 |
| 15 | Fang | 2018 | ApoE-/- | 46.2 | 10.6 | 3 | ApoEWT | 22.1 | 8 | 3 | % | 0.05 | pg | 30305304 | F5A | -1 |
| 16 | Bocan | 1994 | Progression | 0.713 | 0.297 | 8 | Atorvastatin | 0.221 | 0.147 | 8 | mm2 | 0.05/7 | rb | 7840808 | T4 | -1 |
| 17 | Kirk | 1998 | LDLr-/- -IF | 7.39 | 6.7 | 13 | LDLr-/- +IF | 2.01 | 3.4 | 15 | % | 0.05 | ms | 9614153 | F2B | -1 |
| 18 | Niimi | 2013 | Vehicle | 0.173 | 0.15 | 8 | Probucol | 0.015 | 0.0255 | 8 | mm2 | 0.05/6 | rb | 24188322 | F1 | -1 |
| 19 | Lv | 2017 | Control | 0.0386 | 0.04554 | 10 | LDLR-/- | 0.86 | 0.2814 | 10 | mm2 | 0.05/10 | ms | 28983592 | F1D | 1 |
| 20 | Ma | 2012 | Chow | 11904 | 8874.14 | 4 | HFD | 859127 | 210628 | 4 | um2 | 0.05/15 | ms | 22558236 | F2B | 1 |
| 21 | Wang | 2021 | LDLR-/- | 461.5 | 131.9 | 9 | LDLR-/-FbnC1039G+/- | 747.3 | 153.8 | 9 | um2 | 0.05/4 | ms | 33796572 | F4A | 1 |
| 22 | Zhang | 2020 | LDLR-/- | 0.091 | 0.041 | 3 | PM2.5 | 0.219 | 0.056 | 3 | mm2 | 0.05/4 | ms | 31935561 | F2E | 1 |
| 23 | Gao | 2010 | Isotype | 189413 | 99532.7 | 10 | a-IL-17A | 10237 | 35547 | 10 | um2 | 0.05 | ms | 20952673 | F5A | -1 |
| 24 | Engel. | 2019 | WT | 2.2E+05 | 6.5E+04 | 17 | CD47-/- | 3.4E+05 | 1.9E+05 | 17 | um2 | 0 | ms | 31337788 | F1C | 1 |
| 25 | Boriss. | 2013 | ApoE-/-FII-/+ | 9.1E+05 | 3.5E+05 | 10 | ApoE-/- | 1.6E+06 | 6.6E+05 | 10 | um2 | 0.05 | ms | 23409043 | F1B | 1 |
| 26 | Brampt. | 2021 | LDLR-/- | 20.13 | 2.42 | 6 | LDLR-/-Abcc6-/- | 39.3 | 5.22 | 7 | % | 0.05/3 | ms | 33594095 | F1A | 1 |
| 27 | Park | 2016 | NFD | 0.296 | 0.11 | 5 | HFD | 8.33 | 1.52 | 3 | mm2 | 0.05/6 | ms | 26950217 | F1B | 1 |
| 28 | Park | 2016 | LDLR-/- | 640.7 | 175.7 | 10 | LDLR-/-Creb3l3-/- | 854.4 | 157.5 | 10 | um2 | 0.05 | ms | 27417587 | F2C | 1 |

*Abbreviations: see Table S1.*

# Supplementary References